\documentclass[reprint,twocolumn,longbibliography,floatfix,amsmath,amssymb,aps,pra,superscriptaddress]{revtex4-2}

\usepackage{soul}
\usepackage{amsmath}
\usepackage{amssymb}
\usepackage{mathtools}
\usepackage{graphicx}
\usepackage{dcolumn}
\usepackage{bm}
\usepackage{amsmath}
\usepackage{graphicx}
\usepackage{amsfonts}
\usepackage{subfigure}
\usepackage{graphicx}
\usepackage{array}
\usepackage{float}
\usepackage{color}
\usepackage[colorlinks=true,linkcolor=blue]{hyperref}%
\hypersetup{allcolors=blue}
\usepackage[normalem]{ulem}
\usepackage{xcolor}
\usepackage{multirow}

\usepackage{mathtools}
\DeclarePairedDelimiterX\braket[2]{\langle}{\rangle}{#1 \delimsize\vert #2}

\newcommand{\qty}[2]{\ensuremath{#1\,\text{#2}}} 
\newcommand{\PreserveBackslash}[1]{\let\temp=\\#1\let\\=\temp}
\newcolumntype{C}[1]{>{\PreserveBackslash\centering}p{#1}}
\newcommand{\FP}{Fabry-P\'erot }
\newcommand{\Nitro}{N$_2$ }

\begin{document}
\preprint{APS/123-QED}

\title{Rydberg electromagnetically induced transparency of $^{85}$Rb vapor in Ar, Ne and \Nitro gases}
\date{\today }

\author{Bineet Dash}
\email{bkdash@umich.edu}
\affiliation{Department of Physics, University of Michigan, Ann Arbor, MI 48109, USA}
\author{Nithiwadee Thaicharoen}
\email{nithi@umich.edu}
\affiliation{Department of Physics and Materials Science, Faculty of Science, Chiang Mai University, Chiang Mai 50200, Thailand}
\author{Eric Paradis}
\affiliation{Department of Physics \& Astronomy, Eastern Michigan University, Ypsilanti, MI 48197, USA}
\author{Alisher Duspayev}
\affiliation{Department of Physics, University of Michigan, Ann Arbor, MI 48109, USA}
\author{Georg Raithel}
\affiliation{Department of Physics, University of Michigan, Ann Arbor, MI 48109, USA}

\begin{abstract}

An experimental study on Rydberg electromagnetically induced transparency (EIT) in rubidium (Rb) vapor cells containing inert gases at pressures $\le 5$~Torr is reported. Using an inert-gas-free Rb vapor cell as a reference, we measure frequency shift and line broadening of the EIT spectra in Rb vapor cells with argon, neon or nitrogen gases at pressures ranging from a few mTorr to 5~Torr. The results qualitatively agree with a pseudo-potential model that includes $s$-wave scattering between the Rydberg electron and the  inert-gas atoms, and the effect of polarization of the  inert-gas atoms by the Rydberg atoms. Our results are important for establishing Rydberg-EIT as an all-optical and non-intrusive spectroscopic probe for field diagnostics in low-pressure radio-frequency discharges.

\end{abstract}
\maketitle

\section{Introduction}
\label{sec:intro}

Rydberg atoms are widely used for electric field sensing because of their large electric polarizabilities~\cite{gallagher}. Combining optical and microwave excitation of Rydberg states with electromagnetically induced transparency (EIT)~\cite{boller1991} allows spectroscopic measurement of Stark shifts of Rydberg states in response to local electric fields. Recently, Rydberg-EIT has emerged as a broadband, sensitive and SI-traceable field-sensing technique for microwave electrometry~\cite{sedlacek12, holloway14}, radio-frequency (rf) reception ~\cite{receiver2021}, magnetic field~\cite{Ma2017} and dc electric field sensing~\cite{ma2020dc}. Recently, Rydberg atoms have been utilized to measure the macroscopic and microscopic electric fields in cold ion clouds revealing signatures of Holtsmark field distribution and opening an avenue for forthcoming studies on plasma field diagnostics~\cite{Duspayev2023}. Furthermore, the principles of dc electric field sensing using Rydberg states with high angular momentum ($\ell\leq6$) have been demonstrated in room temperature vapor cells~\cite{highellpaper}.

Rydberg-EIT-based techniques are promising candidates for \textit{in-situ} electric field measurement in delicate environments because EIT can enable all-optical signal acquisition. This eliminates or reduces the need for conducting components that must be introduced in the system under test. Such environments include the aforementioned ion clouds and sheath regions of plasma, both of which are characterized by complicated and potentially time-varying electric fields. There is a growing interest in all-optical non-invasive plasma diagnostics methods~\cite{Goldberg_Hoder_Brandenburg_2022} to avoid the limitations of traditional electrostatic probes caused by probe-induced perturbation~\cite{Godyak_2011}.While Stark-effect-based field measurements have been conducted using laser-induced fluorescence spectroscopy of Rydberg states in plasma discharges~\cite{Czarnetzki_1999, shoemaker_1988, Moore_Davis_Gottscho_1984}, this method is limited to hydrogen and other types of low-pressure plasmas. Laser spectroscopy of Rydberg states has been used to measure electric fields in the range of 10~V/cm to 50~V/cm in Coulomb-expanding one-component Rb$^+$ micro-plasmas~\cite{Feldbaum2002}. Other non-invasive field diagnostic methods, such as emission spectroscopy~\cite{brose_oes1,kuraica_oes2} or field-induced non-linear effects \textit{e.g.}, four-wave mixing (E-CARS)~\cite{Lempert_ecars2, Goldberg_2015_ecars2} and second-harmonic generation (E-FISH)~\cite{Goldberg_EFISH_2, Dogariu_2017_EFISH_1} are suitable for fields stronger than \qty{100}{V/cm}. On the other hand, EIT-based field sensing has been demonstrated in laser-generated alkali plasmas by optically interrogating locally embedded alkali Rydberg atoms at the sub-V/cm regime~\cite{anderson17, weller19, Duspayev2023}. 

To extend EIT-based field diagnostics to dc and rf discharges, it is crucial to characterize the effect of gases commonly used to sustain such plasmas on Rydberg-EIT. Rydberg-EIT proceeds through a near-resonant intermediate state and it involves a coherent quantum-interference effect~\cite{Scully1997}. This situation substantially differs from past experiments in which gas-induced Rydberg-line shifts and pressure broadening have been explored using multi-photon spectroscopy methods without resonant intermediate states~\cite{bruce1982collision,weber1982impact, Brillet1980, Thompson1987}. Recent EIT experiments have demonstrated ladder-type EIT in the $5S_{1/2} \leftrightarrow 5P_{1/2} \leftrightarrow 5D_{3/2} $ cascade of Rb~\cite{Sargsyan2010}, as well as Rydberg-EIT in room-temperature vapor cells containing \qty{5}{Torr} of neon~\cite{thaicharoen2024}. In the present paper, we extend earlier work by exploring the effect of the inert gases argon, neon and nitrogen on rubidium Rydberg-EIT over a pressure range from \qty{15}{mTorr} to \qty{5}{Torr}. In the analysis of the experimental data we find that the semi-classical pseudo-potential model reliably predicts the sign and magnitude of corresponding pressure- and gas-species-dependent line shifts and broadening. Our study indicates that Rydberg EIT could be suitable to analyze electric fields in Ar plasmas in the range of tens of mTorr, including inductively coupled plasmas.

The paper is organized as follows. We describe details of the experimental setup and the Rydberg-EIT scheme in Sec.~\ref{sec:setup}. EIT spectra obtained in inert-gas cells and the deduced line shifts and broadening are presented in  Sec.~\ref{sec:results}. In Sec.~\ref{sec:analysis}, the pseudo-potential model for the interactions between a Rydberg atom and an inert-gas atom is presented. We estimate the semi-classical predictions for line shifts and broadening based on $s$-wave scattering length and polarizability values of common inert gases found in the literature. We discuss the validity of the semi-classical model and implications for plasma field diagnostics in Sec.~\ref{sec:discussion}. The paper is concluded in Sec.~\ref{sec:conclusion}.

\section{EXPERIMENTAL SETUP}
\label{sec:setup}

\begin{figure}[htbp]
\centering
\includegraphics[width=\linewidth]{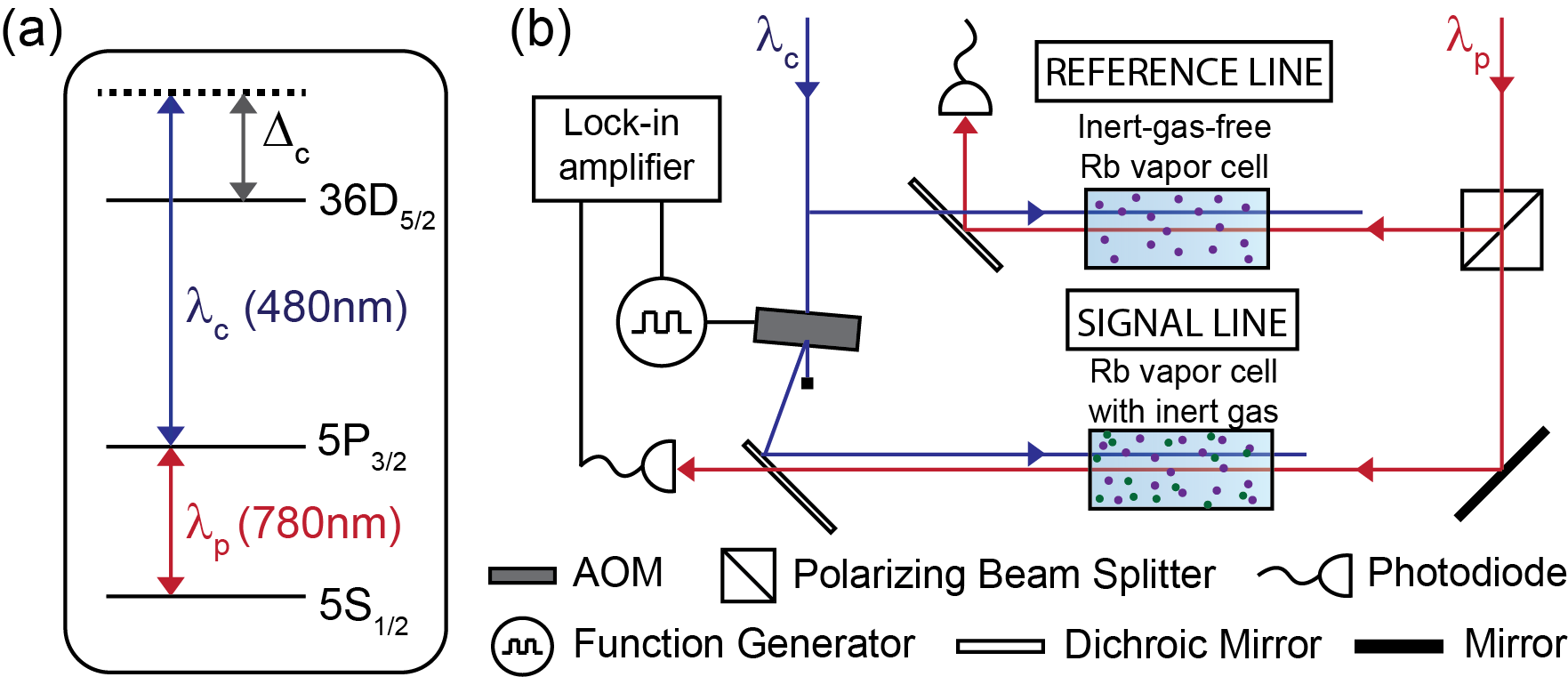}
\caption{(a) Energy levels of $^{85}$Rb used in our experiment. The probe laser ($\lambda_p=$\qty{780}{nm}) is locked to the $5S_{1/2}, F=3 \leftrightarrow 5P_{3/2}, F^\prime = 4$ resonance, and the coupling laser is scanned across the $5P_{3/2} \leftrightarrow 36D_{5/2}$ transition. (b) Experimental setup. (Some elements are omitted for simplicity.)  }
\label{fig:expsetup}
\end{figure}

Our Rydberg-EIT experiments utilize the energy levels of $^{85}$Rb atoms as shown in Fig.~\ref{fig:expsetup}(a). The two-photon scheme uses laser beams of wavelengths $\lambda_p=$\qty{780}{nm} and $\lambda_c=$\qty{480}{nm}, referred to as probe and coupler lasers, respectively. The probe laser is an external-cavity diode laser (ECDL) locked to the $5S_{1/2}, F=3 \leftrightarrow 5P_{3/2}, F^\prime = 4$ transition on the $D_2$ line of $^{85}$Rb using saturation spectroscopy. EIT spectra are obtained by measuring the transmission of the probe power as the coupler-laser frequency is scanned over the $5P_{3/2} \leftrightarrow 36D_{5/2}$ transition. The coupler-laser beam is generated by frequency-doubling the output of another ECDL, which operates at $\sim$~960~nm. The frequency of the coupler laser is calibrated using the transmission of a beam sample of the 960-nm ECDL through a temperature-stabilized \FP (FP) etalon~\cite{hansis2005} with a free spectral range of \qty{250}{MHz}.

The probe and coupler laser beams are split by respective polarizing beam-splitter cubes (PBS) and counter-propagated along two separate paths, each containing a vapor cell (see Fig.~\ref{fig:expsetup}). The polarizations of both beams are linear and parallel in each path. The beams along the upper path in Fig.~\ref{fig:expsetup} (reference line) are aligned through a room temperature inert-gas-free Rb vapor cell of length \qty{7.5}{cm}. There, the probe and coupler beams are Gaussian in shape with respective $1/e^2$ beam waists of \qty{300}{$\mu$m} and \qty{500}{$\mu$m}. The EIT spectra obtained from this cell are used as a reference for the shifts observed in the signal line. The detuning of the coupler laser, $\Delta_c$, is measured relative to the strongest EIT peak in the reference cell (which corresponds to the $5S_{1/2}, F=3 \leftrightarrow 5P_{3/2}, F^\prime=4 \leftrightarrow 36D_{5/2}$ cascade). The transmitted probe power is measured using a fast photo-diode (model - Thorlabs FDS100) and recorded after amplification by a trans-impedance amplifier (TIA) (model - SRS SR570).

The EIT signals from the signal line in Fig.~\ref{fig:expsetup} are recorded in the presence of Ar, Ne or N$_2$. The probe and coupler beams are focused to respective beam waists of \qty{150}{$\mu$m} and \qty{170}{$\mu$m} and counter-propagated through a \qty{5}{cm} vapor cell containing Rb and one of the specified inert gases. 
The optical powers in the signal line are $\approx$ \qty{22}{mW} for the coupler beam and $\approx$ \qty{1.7}{$\mu$W} for the probe beam; the latter corresponds to a Rabi frequency of \qty{30}{MHz}. The coupler beam in the signal line is pulsed at \qty{15}{kHz} at a 50~\% duty cycle using an acousto-optic modulator (AOM) for lock-in detection of the EIT signal, which is required to provide sufficient sensitivity to observe weak EIT lines in cells with higher inert-gas pressures. The EIT-probe power in the signal line is detected using the same types of photo-diode and TIA as in the reference line. The TIA output is sent to a dual-channel lock-in amplifier (MFLI-500, Zurich Instruments), which is referenced to the coupler-beam pulse sequence. The phase of the lock-in amplifier is set by minimizing the $y$-output of the lock-in amplifier on a strong EIT line. The recorded signal is the $x$-output of the lock-in amplifier. 

Using the setup described above, we have obtained EIT signals in 11 inert-gas cells. During data acquisition, the cells are heated to \qty{303}{K}. The TIA in the signal line has a constant gain of \qty{1}{$\mu$A/V}, where the TIA bandwidth is chosen sufficiently high to prevent signal distortion. The EIT spectrum from the reference cell, the EIT spectrum from the cell with inert gas ({\sl{i.e.}}, the $x$-output of the lock-in amplifier), and the transmission of the 960-nm beam sample through the above-described FP etalon are recorded simultaneously to ensure the absence of coupler-laser frequency drifts between the measurements. For each  individual scan, the FP transmission of the 960-nm sample is used to map the coupler-laser scan onto a well-defined frequency axis. The zero-point of the frequency axis is set to the position of the $5S_{1/2}, F=3 \leftrightarrow 5P_{3/2}, F^\prime=4 \leftrightarrow 36D_{5/2}$ EIT line from the reference line. The method eliminates the effects of slow frequency drifts of the coupler laser. 
The spectra shown in the following figures are obtained from taking averages over 200 such scans, for each of the 11 inert-gas Rb cells.

\section{RESULTS}
\label{sec:results}

\begin{figure*}[htb]
	\centering
	\includegraphics[width=\linewidth]{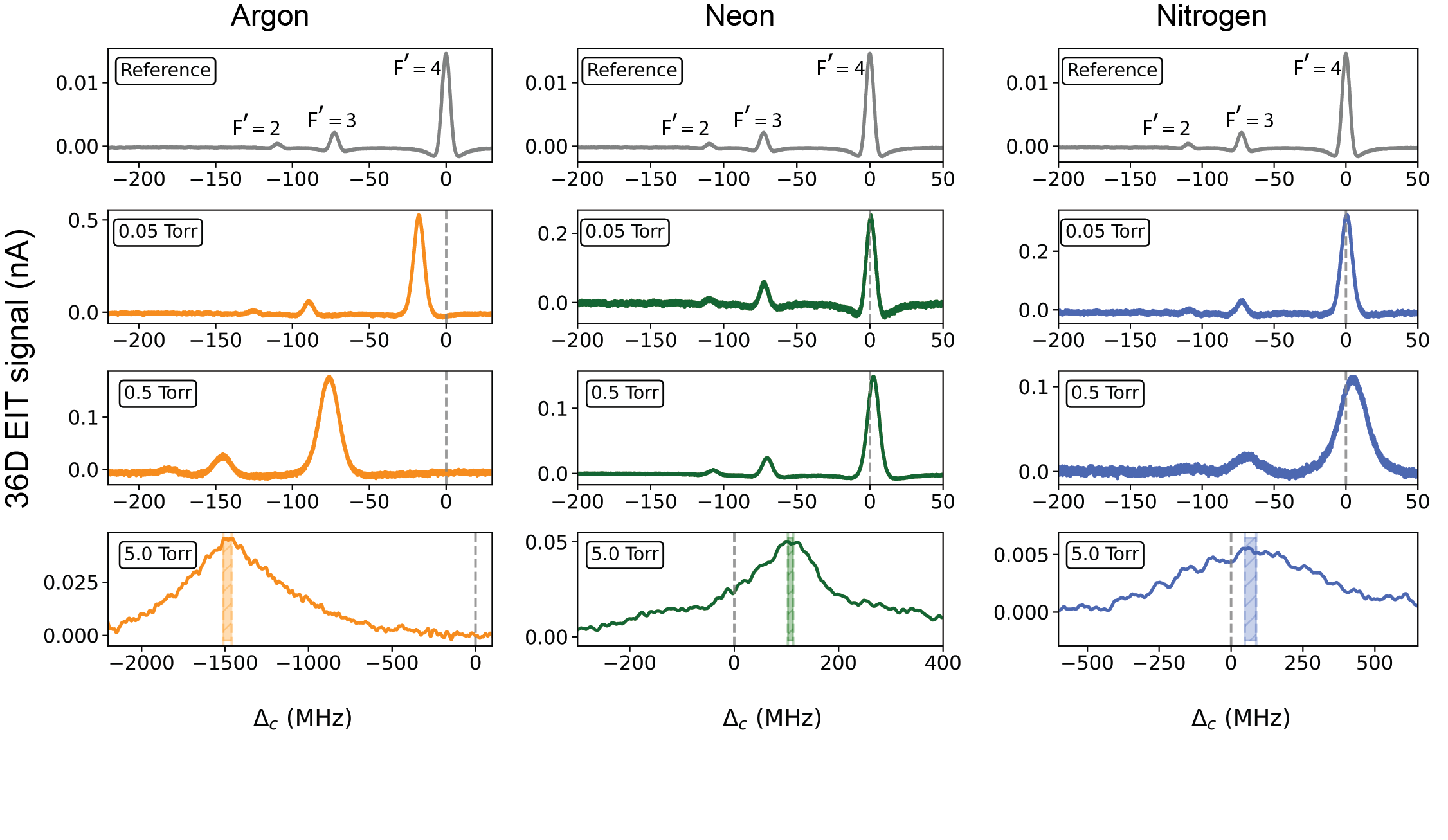}
        \vspace*{-18mm}
	\caption{ Measured EIT spectra.  The top row shows EIT spectra from the inert-gas-free reference cell, while the rows below show spectra for inert-gas-filled cells at the indicated pressures. Left, middle and right columns correspond to cells containing Ar, Ne and N$_2$, respectively. The dashed grey lines represent the locations of the strongest EIT peak in the reference cell. The spectra in the bottom row, obtained from \qty{5}{Torr} cells, are smoothed by a moving average over a \qty{5}{MHz} window. The shaded regions in the bottom row denote the experimental uncertainties in the peak frequency.}
\label{fig:sample_spectra}
\end{figure*}

Selected EIT spectra obtained using inert-gas-filled and inert-gas-free cells are shown in Fig.~\ref{fig:sample_spectra}. Reference EIT spectra obtained from the inert-gas-free vapor cell are shown in the top row in Fig.~\ref{fig:sample_spectra}, where the strongest peak corresponds to the $5S_{1/2}, F=3 \leftrightarrow 5P_{3/2}, F^\prime=4 \leftrightarrow 36D_{5/2} $ EIT line. The smaller peaks to the left correspond to hyperfine sublevels of $5P_{3/2}$, {\sl{i.e.}} $ F^\prime=3 $ and $2$, respectively. The peak frequencies are determined by local parabolic fits. Line shifts at \qty{0.05}{Torr} and \qty{0.5}{Torr} are given by the shifts of the $F'=4$ EIT peaks in the inert-gas-filled cells relative to the $F'=4$ peak positions in the inert-gas-free cell. At \qty{5}{Torr}, the weak $F^\prime=3 $ and $2$ lines become indiscernible. 

\begin{figure}[ht]
\centering
\includegraphics[width=\linewidth]{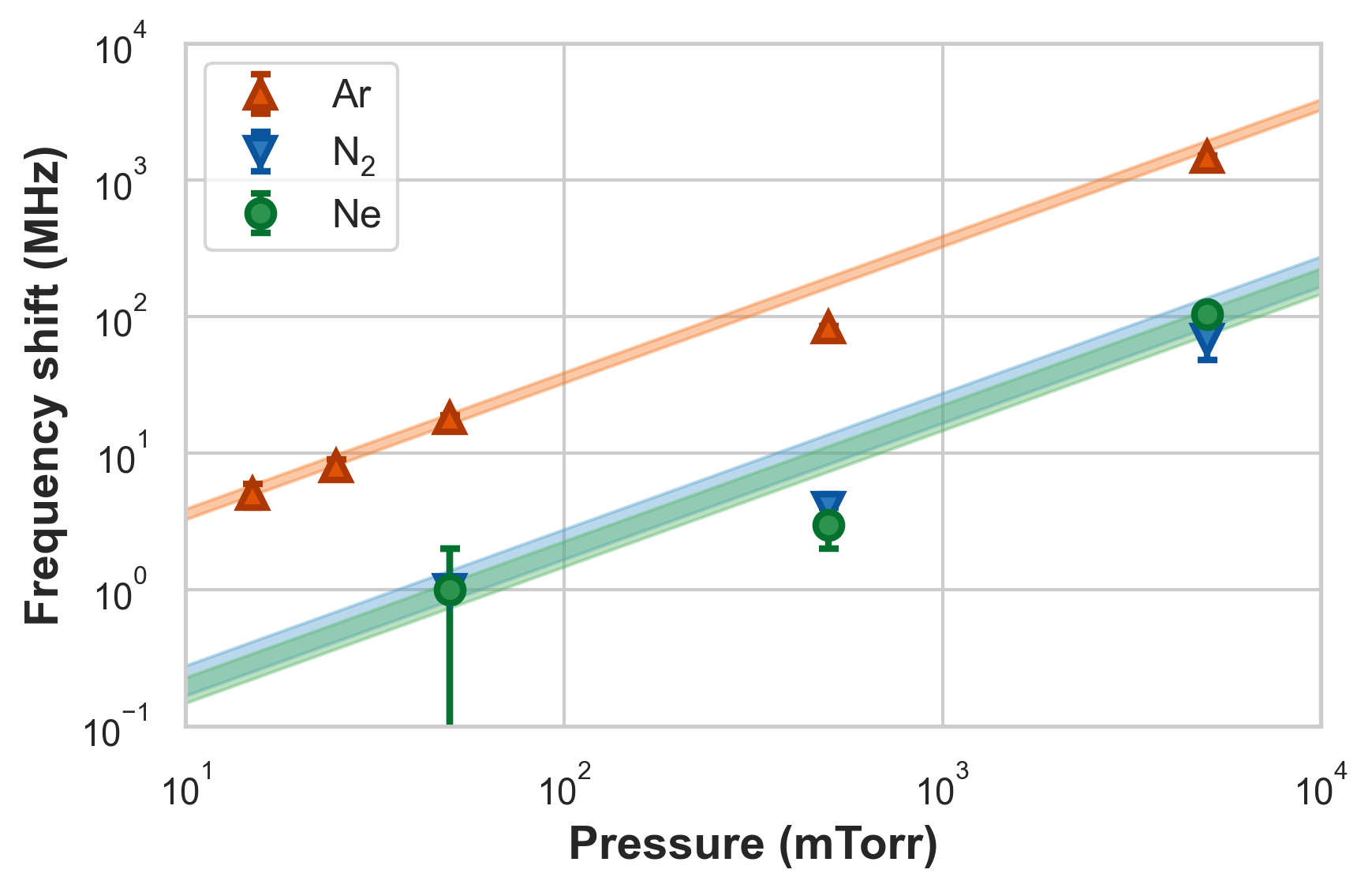}
\caption{Absolute values of the line shifts in the inert-gas-filled cells versus pressure. The shifts are positive for Ne and N$_2$ and negative for Ar. The shaded lines are calculated using Eqs.~\ref{eq:additive-effects} and~\ref{eq:omont-eqs}. The widths of the shaded lines reflect the full variation ranges of the scattering lengths listed in Table~\ref{tab:theory_values}.}
\label{fig:shifts_all}
\end{figure}

\begin{figure}[ht]
\centering
\includegraphics[width=\linewidth]{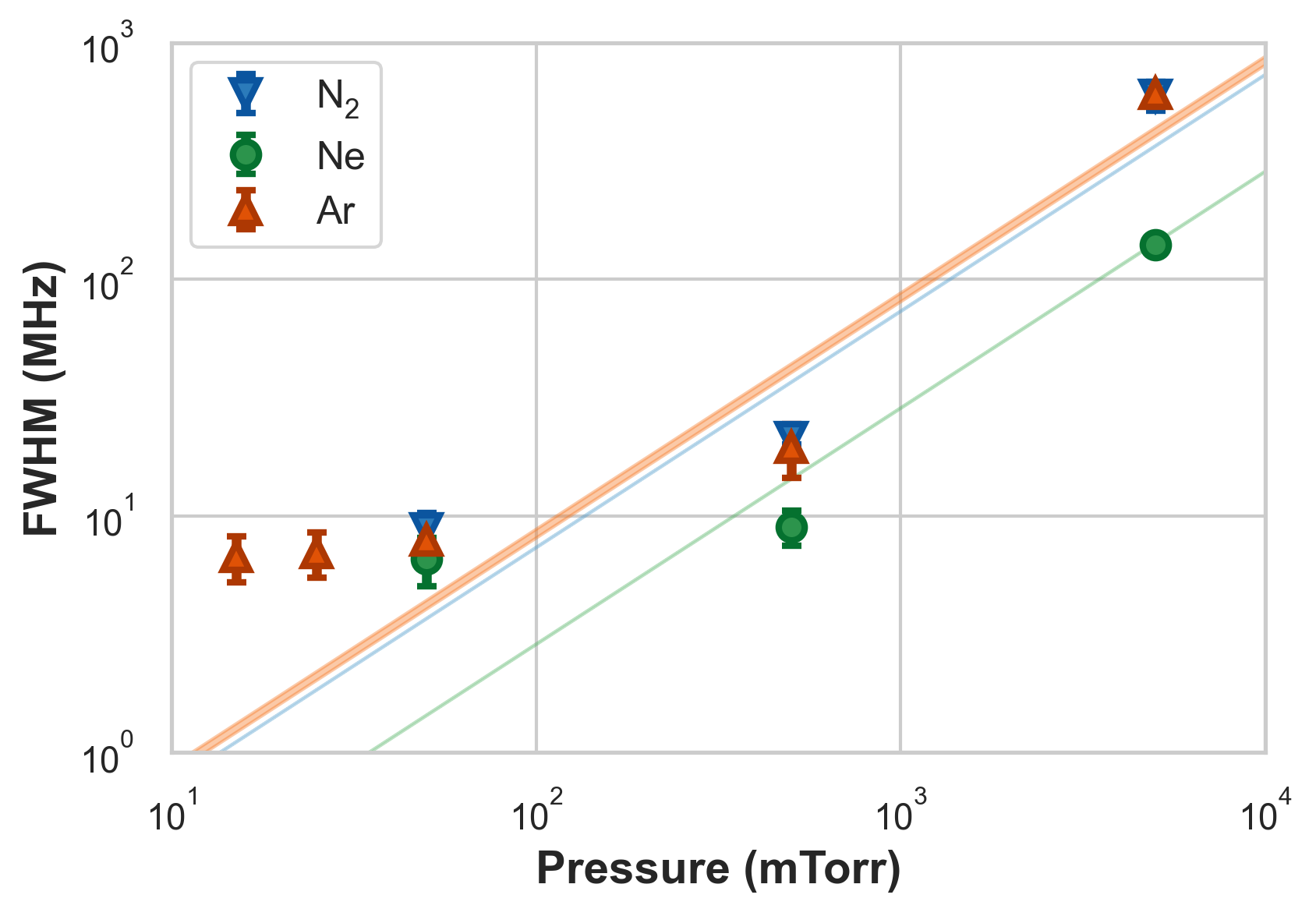}
\caption{Broadening in inert-gas-filled cells versus pressure. The shaded lines are calculated using Eqs.~\ref{eq:additive-effects} and~\ref{eq:omont-eqs}. The widths of the shaded lines reflect the full variation ranges of the scattering lengths listed in Table~\ref{tab:theory_values}. The full widths at half maxima (FWHM) at pressures $\le 50$~mTorr are dominated by inert-gas-free EIT linewidth ($\approx$ \qty{5}{MHz}).}
\label{fig:fwhm_all}
\end{figure}

EIT measurements were made at pressures ranging from \qty{15}{mTorr} to \qty{5}{Torr} of Ar and \qty{50}{mTorr} to \qty{5}{Torr} of Ne and \Nitro. Figs.~\ref{fig:shifts_all} and \ref{fig:fwhm_all} show the shift and the broadening of the Rydberg-EIT lines versus pressure, respectively. As discussed in the next Section, the estimated inert-gas-induced broadening is $\approx$ \qty{1}{MHz} in low pressure cells ({\sl{i.e.}} \qty{50}{mTorr} or smaller), which is about 5 times smaller than the inert-gas-free EIT linewidth and not discernible within our experimental precision. Therefore, we infer the quantitative effect of the inert gases from observations in the higher-pressure cells (\qty{0.5}{Torr} and \qty{5}{Torr}), where shifts and broadening become more significant.

We find that Ar induces a large negative frequency shift, unlike other inert gases considered here. The Rydberg-EIT line is shifted by \qty{-1485 \pm 25}{MHz} and broadened by \qty{610 \pm 30}{MHz} in the \qty{5}{Torr} Ar cell. It should be noted that the log-log plot in Fig.~\ref{fig:shifts_all} shows the absolute value of the frequency shift in Ar. In the \qty{5}{Torr} Ne cell, we observe a moderate positive frequency shift of \qty{105\pm 5}{MHz} and a line broadening of \qty{140\pm 5}{MHz}. The results in the \qty{5}{Torr} Ne cell agree well with earlier findings~\cite{thaicharoen2024} (which were obtained without using lock-in detection).
At 5~Torr of N$_2$ we observe a large line broadening of \qty{600 \pm 85}{MHz}, which contrasts with a rather small positive frequency shift of only \qty{68 \pm 20}{MHz}.

The experimental uncertainties in the line shifts and broadenings in the cells with pressures below \qty{0.5}{Torr} are~$\approx$ \qty{1}{MHz} and are attributed to coupler-laser frequency calibration uncertainty. In the high-pressure cells (\qty{5}{Torr}), the EIT signals are two to three orders of magnitude weaker than in the low-pressure cells, which is due to large inert-gas-induced EIT line broadening. The large linewidths in the high-pressure cells result in larger experimental uncertainties, which are determined by visual inspection of the spectra. An important systematic uncertainty in our experiments arises from the uncertainty of the inert-gas pressures within the cells. The reported pressures are nominal manufacturer values, which have up to \qty{5}{\%} error in the high-pressure cells and \qty{10}{\%} for the cells with pressures below \qty{50}{mTorr}~\cite{pgb}.

\section{Analysis}
\label{sec:analysis} 

Pressure broadening and line shifts of Rydberg levels in low-pressure background gases have been studied extensively since the 1960s using single or two-photon spectroscopy. The collisional shifts of Rydberg transitions were first explained by Fermi using a so-called pseudo-potential~\cite{fermi1934sopra} that treats the ionic core and the Rydberg electron separately in their corresponding interactions with perturbers that can be inert-gas atoms or molecules. The model includes two components: scattering between the perturbers and the Rydberg electron, and a long-range attractive potential experienced by the perturber atoms/molecules due to polarization by the ionic core of the Rydberg atom. A semi-classical treatment is afforded by the impact approximation, in which radiative coupling or change in the Rydberg-electron wave function during the collision are ignored. The scattering is well-approximated by the $s$-wave scattering due to the low kinetic energy of the bound Rydberg electron. 

Refs.~\cite{alekseev1966spectroscopic} and~\cite{Omont1977} have expanded Fermi's treatment by including phase shifts from scattering. In this picture, the frequency shift ($\Delta_r$) and broadening ($\gamma_r$) of a Rydberg line are given by the sum of respective contributions from the scattering and polarization effect; \textit{i.e.}, 
\begin{subequations}
\begin{align}
    \Delta_r &= \Delta_{sc} + \Delta_{p} \\
    \gamma_r &= \gamma_{sc} + \gamma_{p} 
\end{align}
\label{eq:additive-effects}
\end{subequations}
Subscripts $sc$ and $p$ denote the scattering and polarization contributions, respectively, and are expressed as
\begin{subequations}
\begin{align}
     \Delta_{sc} &= a_s N \left[\frac{e^2 a_0}{4 \pi \epsilon_0 \hbar}\right] \\
      \Delta_{p} &= \frac{-6.21 N}{2\pi} \left[\frac{e^2 \alpha}{\left(4 \pi \epsilon_0\right)^2 \hbar}\right]^{2 / 3} v^{1 / 3}\\
      \gamma_{sc} &= 2\times \frac{4 a_s^2 N}{2\pi n}\left[\frac{e^2 a_0^2}{4 \pi \epsilon_0 \hbar}\right] \\ 
      \gamma_{p} &= 2\times \frac{3.59N}{2\pi} \left[\frac{e^2 \alpha}{\left(4 \pi \epsilon_0\right)^2 \hbar}\right]^{2 / 3} v^{1 / 3}
\end{align}
\label{eq:omont-eqs}
\end{subequations}

\noindent where $a_s$ and $\alpha$ denote the $s$-wave electron scattering length and the dipole polarizability of the perturber, respectively, $v$ is the root-mean-square velocity of the perturbers, $n$ is the effective quantum number of the Rydberg state, and $N$ is the volume number density of perturbers. All physical quantities in Eq.~\ref{eq:omont-eqs} are in SI units, including the line shift and the broadening being in units of Hz. The negative sign of $\Delta_{p} $ is due to the attractive polarization potential produced by the ionic core. However, the Rydberg-electron scattering can induce a positive or negative shift, depending on the sign of the scattering length. 

We have surveyed the electric-dipole polarizability, $\alpha$, and the $s$-wave scattering length, $a_s$, of the relevant inert gases in the existing literature. Reported polarizability values from theoretical predictions and several experiments demonstrate excellent agreement with each other. Techniques such as dielectric constant gas thermometry~\cite{fellmuth2018}, optical spectroscopy~\cite{newell1965_polarizability} and capacitance measurement~\cite{buckley2000toroidal} have been used to measure $\alpha$ for common inert gases. The vast majority of results reported in literature for Ar, Ne and N$_2$ agree with each other to within $1\%$. The variation range of the reported $\alpha$-values has a negligible effect on the uncertainties of line shifts and broadenings relevant in our work.

The $s$-wave scattering length is typically estimated by extrapolating experimentally measured electron collision cross-sections to zero energy electrons using modified effective range theory (MERT)~\cite{omalley_1963_mert_original, omalley_1963_mert_applied}. Experimentally reported values of $a_s$ have up to \qty{20}{\%} of variation depending on the collision experiments, \textit{e.g.}, crossed-beam, swarm and time-of-flight measurements, among others (see topical reviews in ~\cite{Dunning_1995_vlereview,crompton1994_swarmreview,brunger2002_vle_moleculesreview}), as well as specifics of the MERT model. As an example, in Table~\ref{tab:ar_database} we show the variation range of experimental values reported for $a_s$ of Ar. In Table~\ref{tab:theory_values} we provide the upper and lower bounds of $a_s$-values reported for Ar, Ne and N$_2$ that we have found. The latter are used for the uncertainty ranges of the calculated shaded lines in Figs.~3 and~4.
 
After accounting for shifts of the ground state and the intermediate state in the utilized EIT energy-level cascade and assuming that the probe laser in on resonance, the final shift of the EIT resonance is given by~\cite{Ma2017}
\begin{equation}
    \Delta = \Delta_{sc} + \Delta_{p} + \left( \frac{\lambda_p}{\lambda_c} -1 \right) \Delta_{e} - \frac{\lambda_p}{\lambda_c} \Delta_{g}
    \label{eq:EITshift}
\end{equation}
where $\Delta_{e} $ and $\Delta_{g} $ denote the absolute energy shifts of the intermediate $5P_{3/2}$ and the $5S_{1/2}$ ground state, respectively. 
It is seen that the EIT shift depends independently upon $\Delta_{g}$, $\Delta_{e}$ and $\Delta_r = \Delta_{sc} +\Delta_p$, with different scaling factors.
Notably, collisional shifts of the $D_2$ line, as commonly reported in literature~\cite{rotondaro1997collisional,Sargsyan2010,Ottinger1975}, only provide $\Delta_{e} -\Delta_{g}$, which is not sufficient to evaluate Eq.~\ref{eq:EITshift}. Under the assumption that the collisional shift of the ground state $5S_{1/2}$ is smaller than that of the excited electronic state $5P_{3/2}$, one may conclude that $\Delta_e \approx \Delta_{D_2}$, the shift of the $D_2$ line. The inert-gas-induced shift of the $D_2$ line is $\approx$ \qty{2}{MHz/Torr} for Ne~\cite{Ottinger1975} and $\approx$ \qty{6}{MHz/Torr} for Ar and \Nitro~\cite{rotondaro1997collisional}. The line broadening rates are $\approx$ \qty{10}{MHz/Torr} for Ne and $\approx$ \qty{20}{MHz/Torr} for Ar and \Nitro~\cite{Sargsyan2010,rotondaro1997collisional, Ottinger1975}.These shifts and broadenings are a factor of three or more smaller than the respective Rydberg energy shifts and broadenings calculated from Eq.~\ref{eq:omont-eqs}. Therefore, we neglect the inert-gas effects on the $D_2$ line at our current level of experimental precision. With this assumption, we attribute the Rydberg-EIT line shift and broadening solely to the Rydberg state, \textit{i.e.}, $\Delta \approx \Delta_r$ and $\gamma \approx \gamma_r$. 

Table~\ref{tab:theory_values} summarizes calculated polarization and scattering contributions, $\Delta_p$ and $\Delta_{sc}$, respectively, and $\Delta_r = \Delta_p + \Delta_{sc}$, as well as corresponding broadenings, for the inert gases studied in our work, in units MHz/Torr at 303~K. Eq.~\ref{eq:omont-eqs} reveals the linear dependence of line shift and broadening on inert gas pressure, which is an important result of the semi-classical model. We find in Fig.~\ref{fig:shifts_all} that frequency shift scales linearly with inert-gas pressures and agrees with estimates from Eq.~\ref{eq:omont-eqs} within the uncertainty of our measurement.

\begin{table}[hbt]
\centering
        \begin{tabular}{cc}
            \hline
           $a_s$  & \multirow{2}{*}{Reference}  \\
            (in au) &  \\ \hline 
           -1.50  & Milloy et. al. (1977)~\cite{milloy1977momentum}\\
           -1.488 & Haddad and O’Malley (1982)~\cite{haddad1982scattering} \\
           -1.506 & McEachran and Stauffer (1983)~\cite{mceachran1983elastic} \\
           -1.449 & Ferch et al. (1985)~\cite{ferch1985electron} \\
           -1.492 & Buckman and Lohmann (1986)~\cite{buckman1986low} \\
           -1.641 & Weyhreter et al. (1988)~\cite{weyhreter1988measurements} \\
           -1.686  & Pack et al. (1992)~\cite{pack1992longitudinal}  \\
           -1.442 & Buckman and Mitroy (1989) ~\cite{buckmanmittroy1989}  \\
           -1.459 & Petrovi\'{c} et al. (1995)~\cite{petrovic1995ar} \\
           -1.365 & Kurokawa et. al. (2011)~\cite{kurokawa2011high} \\
        \hline
        \end{tabular}
    \caption{Survey of s-wave scattering length ($a_s$) of Ar reported from past experiments.}
    \label{tab:ar_database}
\end{table}

\begin{table*}[ht]
    \centering
    \begin{tabular}{C{1cm}  C{5cm}  C{3cm} C{1.25cm}C{1.25cm}C{1.25cm}C{1.25cm}C{1.25cm}C{1.25cm}}
         \hline
        \multirow{2}{*}{Species} &  Range of $a_s$ & $\alpha$ & $\Delta_{sc}$ &  $\Delta_{p}$ & $\Delta_r$ & $\gamma_{sc}$ & $\gamma_{p}$ & $\gamma_r$ \\
        &  (au) & (au) & \multicolumn{3}{c}{(MHz/Torr)} & \multicolumn{3}{c}{(MHz/Torr)} \\
        \hline     
        Ar & -1.365 to -1.69~\cite{milloy1977momentum, haddad1982scattering,ferch1985electron,buckman1986low,weyhreter1988measurements,pack1992longitudinal, buckmanmittroy1989, petrovic1995ar, kurokawa2011high} & 11.07~\cite{orcutt1967dielectric, fellmuth2018, buckley2000toroidal, newell1965_polarizability} & -290.9 & -58.1 & -349.0 & 15.9 & 67.1 & 83.0 \\
        Ne &   0.20 to 0.24~\cite{sol_1975_ne, Gulley_1994_ne,Hoffmann1969_ne,Koizumi_1984_ne, OMalley_1980_ne,buckmanmittroy1989, stauffer_1985_ne,Robertson_1972_ne}  & 2.66~\cite{orcutt1967dielectric, Gaiser_2010, fellmuth2018} & 41.8 & -24.5 & 17.3  & 0.3 & 28.2 &  28.5 \\
        \Nitro & 0.404 to 0.460~\cite{idziaszek2009modified,fabrikant1984effective, morrison_1997_n2, Chang_1981_n2} &  11.74~\cite{newell1965_polarizability,orcutt1967dielectric} & 78.9 & -63.1 &  15.8  & 1.2 & 72.8 & 74.0 \\
        \hline
    \end{tabular}
    \caption{Line shifts and broadenings based on pseudo-potential model calculated using Eq.~\ref{eq:omont-eqs}. For $a_s$ we show the upper and lower bounds of the results from the references cited. The values shown for $\alpha$ are consistent with the references cited, at the level of precision given in our table. In our calculation of the $\Delta_*$ and $\gamma_*$, we use $a_s= -1.49 a_0$, $0.214 a_0$ and $0.404 a_0$ for Ar, Ne and N$_2$, respectively, as well as the $\alpha$-values shown.}
    \label{tab:theory_values}
\end{table*}

\section{Discussion}
\label{sec:discussion}

The presented semi-classical model rests on the validity of impact approximation and the assumption that the Rydberg electron and the Rydberg-atom ionic core are decoupled in their respective interaction with the perturbers. We emphasize that both of these assumptions are applicable in our experimental regime of highly excited Rydberg levels and low inert-gas pressure, as discussed below. 

Semi-classical treatments of the pseudo-potential model ~\cite{kaulakys1984broadening, henry2002collisional} have established that the broadening cross-section reaches an asymptotic value for effective quantum numbers exceeding $n_{\text{max}} \approx |a_s|^{1/3}/(\alpha v^5)^{1/18} $ ($a_s$, $\alpha$ and $v$ in atomic units). The broadening cross-section at higher Rydberg levels arises from the polarization potential of the ionic core, whereas the effect of the Rydberg electron is limited to the scattering described by the Fermi interaction. For the inert gases studied in our work, it is $n_{\text{max}} \approx 10$, whereas our experiment is performed with $36D_J$ states. Therefore, the use of Eq.~\ref{eq:additive-effects} is well-justified, {\sl{i.e.}}  the components from long-range polarization potential and electron scattering can be treated separately. 

Next, we consider the validity of the impact approximation, which holds when the timescale of collisions is much faster than the mean-free time, or, equivalently, when the mean number of perturbers occupying the interaction volume is much smaller than unity. This limits the perturber density to $ N \ll 3v/(\pi^2 \alpha)$ (in atomic units) ~\cite{kaulakys1984broadening}, corresponding to $\approx$ \qty{100}{Torr} of inert-gas pressure. This is considerably larger than the highest pressures utilized in our experiments. 

The validity of the semi-classical treatment is highlighted by the observed linear scaling of the line shift and broadening with inert-gas pressure at sufficiently high inert-gas pressure. Other approximations that are implicit in Eq.~\ref{eq:omont-eqs} include the $s$-wave treatment of the Fermi interaction and the assumption of negligible inelastic scattering. While the former is generally justified for low-energy Rydberg electrons, the latter is more suitable for the interaction of atomic perturbers like Ne and Ar with Rydberg atoms, but less so for molecules. Non-polar molecular perturbers like N$_2$ possess a quadrupole moment and a rich rovibrational structure, which can give rise to several inelastic scattering processes~\cite{BEIGMAN1995}. The study of such additional cross-sections in the context of Rydberg-EIT is beyond the scope of our present paper. 

Our results have important implications for future applications of Rydberg-EIT to plasma field diagnostics. The gases studied in our experiment are common choices for low-pressure plasma discharges, including inductively coupled and dc plasmas. Rf discharges in gases at pressures above \qty{1}{mTorr} rely on power absorption through collisional heating. This mechanism is most efficient at gas pressures at which the effective rate of collisions between electrons and neutral atoms is close to the angular frequency of the rf power applied, \textit{i.e.}, $ \nu_{en} \approx 2\pi f_{\text{rf}}$.  For instance, Ar offers efficient rf power absorption at the standard frequency of \qty{13.56}{MHz} over a range of pressures from \qty{10}{mTorr} to \qty{100}{mTorr} due to favorable values of $ \nu_{en} $~\cite{Godyak_1994,kralkina2016rf}. In this regime, the gas-induced EIT broadening is negligible in comparison to typical EIT linewidths, as shown in Fig.~\ref{fig:shifts_all}, and the estimated line shift is less than \qty{20}{MHz} for Ar and \qty{2}{MHz} for Ne and N$_2$. This indicates the viability of Rydberg-EIT as a diagnostic tool for electric fields in low-temperature rf plasma generated using the gases utilized in our study. 

\section{Conclusion}
\label{sec:conclusion}

We have demonstrated that Rydberg-EIT presents  a viable spectroscopic method in the presence of the commonly used inert gases Ar, Ne and N$_2$ at pressures up to about \qty{5}{Torr}. Estimates based on the semi-classical pseudo-potential model of interactions between a Rydberg atom and an inert-gas atom agree well with our experimentally observed EIT line shifts and broadenings, within the achieved uncertainty. In the sub-\qty{100}{mTorr} pressure regime, inert-gas-induced broadening is negligible compared to the typical linewidth of our Rydberg-EIT signals. This pressure range is frequently used for low-temperature rf plasma discharges. Considering EIT shift, broadening and signal strength, we believe that Ar will be a particularly good choice in future plasma-physics applications of Rydberg-EIT.

\section*{Acknowledgments}
\label{sec:acknowledgments}

We thank Dr. Ryan Cardman for early work on this experiment and Dr. David A. Anderson from Rydberg Technologies Inc. for valuable discussions. This project was supported by the U.S. Department of Energy, Office of Science, Office of Fusion Energy Sciences under award number DE-SC0023090. N.T. acknowledges funding from Fundamental Fund 2023, Chiang Mai University. A.D. acknowledges support from the Rackham Predoctoral Fellowship at the University of Michigan.

\bibliography{references.bib}

\end{document}